\documentclass[aps,amsmath,twocolumn,prb,showpacs]{revtex4}

\usepackage{graphicx}
\usepackage{color}

\begin{document}


\title{Thermodynamics and linear response of a Bose-Einstein condensate of microcavity polaritons}

\author{Davide Sarchi and Vincenzo Savona}

\affiliation{Institute of Theoretical Physics, Ecole Polytechnique
F\'ed\'erale de Lausanne EPFL, CH-1015 Lausanne, Switzerland}
\email[]{davide.sarchi@epfl.ch}


\begin{abstract}
In this work we derive a theory of polariton
condensation based on the theory of interacting Bose particles. In
particular, we describe self-consistently the linear
exciton-photon coupling and the exciton-nonlinearities, by
generalizing the Hartree-Fock-Popov description of BEC to the
case of two coupled Bose fields at thermal equilibrium. In this
way, we compute the density-dependent one-particle spectrum, the
energy occupations and the phase diagram. The results
quantitatively agree with the existing experimental findings. We
then present the equations for the linear response of a polariton
condensate and we predict the spectral response of the system to
external optical or mechanical perturbations.
\end{abstract}

\maketitle

\section{Introduction}

An important advance in the investigation of quantum fluids was
recently achieved with the experimental observation of high
quantum degeneracy and off-diagonal long-range coherence, in a
gas of exciton- polaritons in a two-dimensional semiconductor
microcavity \cite{kasprzak06,balili07}. High quantum degeneracy
has been also observed in long-living polariton systems close to
thermal equilibrium \cite{deng06}. Bose-Einstein condensation
(BEC) is the most natural way of describing these findings.
However, due to the peculiarity of the polariton system, in
particular the finite polariton lifetime, the intrinsic 2-D
nature and the presence of interface disorder, the existing
theoretical frameworks rather interpret the phenomenon in strict
analogy either with laser physics \cite{laussy04,schwendimann06}
or with the BCS transition of Fermi particles
\cite{keeling04,marchetti06,szymanska06}. In particular, the
problem of the polariton kinetics and of the non-equilibrium
effects, due to the finite polariton lifetime and the relaxation
bottleneck, have been investigated in many works
\cite{schwendimann06,szymanska06,doan05,sarchi06,wouters07}. From
these studies, the main effect of non-equilibrium seems to be a
significant depletion of the condensate - with corresponding loss
of long-range coherence - \cite{sarchi07} and the appearance of a
diffusive excitation spectrum at low momenta
\cite{szymanska06,wouters07}.

In spite of the high relevance of these theoretical descriptions,
two basic questions remain still unanswered. Are the experimental
findings correctly interpreted in terms of a quantum field theory
of interacting bosons? Could the achievement of polariton BEC give
new insights into the fundamental physics of interacting Bose
systems?

In this work we tackle these two questions, by showing that
polaritons can be modeled borrowing from the theory of
interacting Bose particles. In particular, we describe
self-consistently the linear exciton-photon coupling and the
exciton-nonlinearities, by generalizing the Hartree-Fock-Popov
(HFP) description of BEC to the case of two coupled Bose fields at
thermal equilibrium. In this way, we compute the
density-dependent energy shifts and the phase diagram and we find
a very good agreement with the recent experimental findings. Then,
we apply the present theory to derive the full set of equations
describing the density-density response of the polariton
condensate to an external perturbation. Focusing on the photon
density response, which is directly related to photoluminescence,
we predict different response of the collective modes to optical
(light) or mechanical (coherent phonons) perturbations. Since
this behavior is driven by the presence of a coherent exciton
field, we suggest that an experiment investigating these features
could possibly solve the tricky problem of assessing the nature of
the polariton condensate.

\section{Theory}
\label{sec:theory}

The physics of the polariton system is basically that of two
linearly coupled oscillators, the exciton and the cavity photon
fields \cite{savona95,savona99b}. Considering the limit of low
density, the exciton field can be treated as a Bose field,
subject to two kinds of interactions, the mutual exciton-exciton
interaction and the effective exciton-photon interaction,
originating from the saturation of the exciton oscillator
strength \cite{laikhtman07}. Therefore, to describe polariton BEC
we extend to the case of two coupled interacting Bose fields the
formalism adopted in describing the BEC of a single Bose field
\cite{shi98,pita03}.

We express the exciton and photon field operators in the
Heisenberg representation via the notation
\begin{equation}
\hat{\Psi}_x({\bf r},t)=\frac{1}{\sqrt{A}}\sum_{\bf k} e^{i{\bf
k}\cdot {\bf r}} \hat{b}_{\bf k}(t)\,, \label{eq:fieldopb}
\end{equation}
and
\begin{equation}
\hat{\Psi}_c({\bf r},t)=\frac{1}{\sqrt{A}}\sum_{\bf k} e^{i{\bf
k}\cdot {\bf r}} \hat{c}_{\bf k}(t)\,, \label{eq:fieldopc}
\end{equation}
where $A$ is the system area, while $\hat{b}_{\bf k}$ and
$\hat{c}_{\bf k}$ are independent Bose operators $([\hat{b}_{\bf
k},\hat{c}_{\bf k'}^{\dagger}]=0)$. Notice that in this work we
assume scalar exciton and photon fields. However, the theory can
be generalized to include their vector nature, accounting for
light polarization and exciton spin\footnote{Shelykh {\em et al.}
\cite{shelykh06} have recently studied the effects of
polarization and spin at $T=0$, within the Gross-Pitaevskii limit
restricted to the lower polariton field}. We adopt a finite
system area $A$ in order to model the effect of confinement, due
to both the intrinsic disorder \cite{langbein02,richard05b} and
to the finite size of the excitation spot
\cite{deng03,richard05}. While in 2-D, in the thermodynamic
limit, the occurrence of BEC would be prevented by the divergence
of thermal fluctuations \cite{hohenberg67}, the finite size
modifies the density of states, resulting in a finite amount of
thermal fluctuations \cite{lauwers03}. The dependence of the
results on $A$ is discussed in Section \ref{sec:discuss}.

The exciton-photon Hamiltonian, including the exciton
non-linearities, reads
\begin{equation}
\hat{H}=\hat{H}_{0}+\hat{H}_{R}+\hat{H}_{x}+\hat{H}_{s}\,.
\label{eq:Hcomp}
\end{equation}
The non-interacting exciton-photon Hamiltonian is
\begin{equation}
\hat{H}_{0}=\sum_{\bf k}\left(\epsilon^{x}_{\bf k}
\hat{b}^{\dagger}_{\bf k}\hat{b}_{\bf k}+\epsilon^{c}_{\bf k}
\hat{c}^{\dagger}_{\bf k}\hat{c}_{\bf k}\right)\,,
\label{eq:nonintH}
\end{equation}
$\epsilon^{x}_{\bf k}=\hbar^2k^2/2m_x$ is the free exciton energy
dispersion, $m_x$ is the exciton effective mass,
$\epsilon^{c}_{\bf k}=\epsilon^{c}_{\bf 0}\sqrt{1+(k/k_z)^2}$ is
the free photon dispersion, $\epsilon^{c}_{\bf 0}=\hbar (c/n_c)
k_z$, $c$ is the velocity of light, $n_c$ is the refractive index,
$k_z=\pi/L_c$, and $L_c$ is the cavity length. The term
\begin{equation}
\hat{H}_{R}=\hbar\Omega_{R}\sum_{\bf k}(b^{\dagger}_{\bf
k}\hat{c}_{\bf k}+h.c.)\, \label{eq:lincoupH}
\end{equation}
describes the linear exciton-photon coupling. The term
\begin{equation}
\hat{H}_{x}=\frac{1}{2A}\sum_{{\bf k}, {\bf k'}, {\bf
q}}v_{x}({\bf k}, {\bf k'}, {\bf q})\hat{b}^{\dag}_{{\bf k}+{\bf
q}}\hat{b}^{\dag}_{{\bf k'}-{\bf q}}\hat{b}_{{\bf k'}}\hat{b}_{\bf
k}\, \label{eq:XXintH}
\end{equation}
is the effective exciton-exciton scattering Hamiltonian, modeling
both Coulomb interaction and the non-linearity due to the Pauli
exclusion principle for electrons and holes forming the exciton.
The remaining term
\begin{equation}
\hat{H}_{s}=\frac{1}{A}\sum_{{\bf k}, {\bf k'}, {\bf q}}v_{s}({\bf
k}, {\bf k'}, {\bf q})(\hat{c}^{\dag}_{{\bf k}+{\bf
q}}\hat{b}^{\dag}_{{\bf k'}-{\bf q}}\hat{b}_{{\bf k'}}\hat{b}_{\bf
k}+h.c.)\, \label{eq:satintH}
\end{equation}
models the effect of Pauli exclusion on the exciton oscillator
strength \cite{laikhtman07}, that is reduced for increasing
exciton density \cite{schmitt-rink85}. In this work we account
for the full momentum dependence of $v_{x}({\bf k},{\bf k'},{\bf
q})$ and $v_{s}({\bf k},{\bf k'},{\bf q})$
\cite{rochat00,okumura01}. In particular, these potentials vanish
at large momentum, thus preventing the ultraviolet divergence
typical of a contact potential \cite{pita03}, without introducing
an arbitrary cutoff.

\subsection{Bogoliubov ansatz}

To describe the condensed system, we extend the Bogoliubov ansatz
\cite{pita03} to both the exciton and photon Bose fields,
\begin{equation}
\hat{\Psi}_{x(c)}({\bf r},t)=\Phi_{x(c)}({\bf
r},t)+\tilde{\psi}_{x(c)}({\bf r},t)\,, \label{eq:bogans}
\end{equation}
i.e. the total field is expressed as the sum of a classical
symmetry-breaking term $\Phi_{x(c)}$ for the condensate wave
function, and of a quantum fluctuation field
$\tilde{\psi}_{x(c)}$. The Bogoliubov ansatz imposes to introduce
anomalous propagators for the excited particles, describing
processes where a pair of particles is scattered inwards or
outwards the condensate \cite{shi98,fetter71}. The resulting $16$
thermal propagators in the matrix form (in the energy-momentum
representation, assuming a spatially uniform system) are
\begin{equation}
G({\bf k},i\omega_n)=\left(\begin{array}{c c} g^{xx}({\bf
k},i\omega_n) & g^{xc}({\bf
k},i\omega_n) \\
g^{cx}({\bf k},i\omega_n) & g^{cc}({\bf k},i\omega_n)
\end{array}\right), \label{eq:defgreen}
\end{equation}
where the elements of each 2 $\times$ 2 matrix block are
($j,l=1,2$; $\chi,\xi=x,c$)
\begin{equation}
g_{jl}^{\chi\xi}({\bf k},i\omega_n)=-\int_{0}^{\beta}d\tau
e^{i\omega_{n}\tau}\langle \hat{O}_{\chi}^{j}\left({\bf
k},\tau\right)\hat{O}_{\xi}^{l}\left({\bf
k},0\right)^{\dagger}\rangle_{\tau,\beta}\,, \label{eq:gij}
\end{equation}
$\hbar\omega_n=2\pi n/\beta,n=0,\pm 1,...$ are the Matsubara
energies for bosons, $\beta=1/k_BT$ and the symbol $\langle
...\rangle_{\tau,\beta}$ indicates the thermal average of the
time ordered product. Here, to represent the exciton and the
photon fields, we adopt the compact notation
$\hat{O}_{\xi}^{1}({\bf k})=\hat{O}_{\xi}({\bf k})$,
$\hat{O}_{\xi}^{2}({\bf k})=\hat{O}_{\xi}^{\dagger}(-{\bf k})$
and $\hat{O}_{x}({\bf k})=\hat{b}_{\bf k}$, $\hat{O}_{c}({\bf
k})=\hat{c}_{\bf k}$. Correspondingly, the generalized
one-particle density
\begin{equation}
n^{\chi\xi}=n_0^{\chi\xi}+\tilde{n}^{\chi\xi}\,,
\label{eq:densgener}
\end{equation}
with $\chi,\xi=x,c$, is separated into the contribution of the
condensate $n^{\chi\xi}_0=\Phi^{*}_{\chi}\Phi_{\xi}$ and of the
excited particles
\begin{equation}
\tilde{n}^{\chi\xi}=\sum_{\bf k\neq 0}n^{\chi\xi}_{\bf k}= \sum_{\bf
k\neq 0}\langle \hat{O}_{\chi}^{2}({\bf k}) \hat{O}_{\xi}^{1}({\bf
k})\rangle\,. \label{eq:excdensgen}
\end{equation}
This latter quantity represents the excited-state density matrix,
expressed in the exciton-photon basis, and it is directly related
to the corresponding normal propagator via the well known relation
\cite{shi98}
\begin{equation}
\tilde{n}^{\chi\xi}_{\bf k}=-\int \frac{d\omega}{\pi}
\mbox{Im}\{(g^{\chi\xi}_{11})^{ret}({\bf k},\omega)\}n_B(\omega)\,,
\label{eq:densgr}
\end{equation}
where the retarded Green's function
\begin{equation}
(g^{\chi\xi}_{11})^{ret}({\bf k},\omega)=g^{\chi\xi}_{11}({\bf
k},i\omega_n\rightarrow \omega+i0^+)
\end{equation}
is the analytical continuation to the real axis of the
imaginary-frequency Green's function \cite{shi98}.

\subsection{Condensate wave function}

Within the Popov approximation, for a uniform system, the two
coupled equations for the condensate amplitudes are
\begin{eqnarray}
& i\hbar\dot{\Phi}_{x}&=[\epsilon^{x}_0 -2
\mbox{Re}\{\bar{v}_{s}({\bf 0,0})n^{xc}_{\bf 0}+ 2\left.\sum_{\bf
k}\right.^{\prime}
\bar{v}_{s}({\bf k,0})\tilde{n}^{xc}_{\bf k}\} \nonumber \\
&&+(\bar{v}_x({\bf 0,0})n^{xx}_{\bf 0}+2\left.\sum_{\bf
k}\right.^{\prime} \bar{v}_x({\bf k,0})\tilde{n}^{xx}_{\bf k})]\Phi_{x}\nonumber \\
&&+(\hbar\Omega_{R}-\sum_{\bf k} \bar{v}_s({\bf k,0})n^{xx}_{\bf k})\Phi_{c}, \nonumber \\
&i\hbar\dot{\Phi}_{c}&=\epsilon^{c}_0 \Phi_c
+\hbar\Omega_{R}\Phi_{x}\nonumber
\\
&&-[\bar{v}_{s}({\bf 0,0})n^{xx}_{\bf 0}+2\left.\sum_{\bf
k,0}\right.^{\prime} \bar{v}_{s}({\bf k,0})\tilde{n}^{xx}_{\bf
k}]\Phi_{x} \label{eq:GPeq}
\end{eqnarray}
where $\left.\sum_{\bf k}\right.^{\prime}=\sum_{{\bf k}\neq 0}$
and
\begin{equation}
\bar{v}_{x(s)}({\bf k,q})=\frac{1}{2}\left[v_{x(s)}({\bf
k,q,0})+v_{x(s)}({\bf k,q,k-q})\right]\,.
\end{equation}
We assume that both the condensate fields evolve with the same
characteristic frequency $E/\hbar$, i.e.
\begin{equation}
\Phi_{x(c)}(t)=e^{-i \frac{E}{\hbar}t}\Phi_{x(c)}(0)\,.
\end{equation}
By replacing this evolution into Eq. (\ref{eq:GPeq}), we obtain a
generalized set of two coupled time-independent Gross-Pitaevskii
equations, which can be formally written in the matrix form
\begin{equation}
E\left(\begin{array} {c}
X_0 \\
C_0 \end{array}\right)=\hat{L}^{GP}\left(\begin{array} {c}
X_0 \\
C_0 \end{array}\right)\,,\label{eq:GPeq_matr}
\end{equation}
where
\begin{eqnarray}
\hat{L}^{GP}_{11}&=&\epsilon^{x}_0-2 \mbox{Re}\{\bar{v}_{s}({\bf
0,0})n^{xc}_{\bf 0}+ 2\left.\sum_{\bf
k}\right.^{\prime} \bar{v}_{s}({\bf k,0})\tilde{n}^{xc}_{\bf k}\} \\
&&+(\bar{v}_x({\bf 0,0})n^{xx}_{\bf 0}+2\left.\sum_{\bf
k}\right.^{\prime} \bar{v}_x({\bf k,0})\tilde{n}^{xx}_{\bf k}) \nonumber \\
\hat{L}^{GP}_{12}&=&\hbar\Omega_{R}-\sum_{\bf k} \bar{v}_s({\bf k,0})n^{xx}_{\bf k} \nonumber \\
\hat{L}^{GP}_{21}&=&\hbar\Omega_{R}-\bar{v}_{s}({\bf
0,0})n^{xx}_{\bf 0}-2\left.\sum_{\bf k}\right.^{\prime}
\bar{v}_{s}({\bf k,0})\tilde{n}^{xx}_{\bf k}
\nonumber \\
\hat{L}^{GP}_{22}&=&\epsilon^{c}_0\,, \label{eq:GPeq_matr2}
\end{eqnarray}
we have defined the normalized Hopfield coefficients of the
condensate state as $\Phi_{x}=X_0 \Phi$ and $\Phi_{c}=C_0 \Phi$,
satisfying $|X_0|^2+|C_0|^2=1$, and $n_0=|\Phi|^2$ is the actual
density of the polariton condensate. The two solutions
$E=E^{lp(up)}$ of Eq. (\ref{eq:GPeq_matr}) define the lower and
upper polariton condensate modes
\begin{equation}
\Phi_{lp(up)}=X^{lp(up)*}_0{\Phi}_{x}+C^{lp(up)*}_0{\Phi}_{c}\,.
\end{equation}
The condensate energy is given by the lower energy solution
$E_0^{lp}$, which corresponds to the minimal energy of the
polariton states. In the present U(1) symmetry-breaking approach,
we can identify the condensate energy with the chemical potential
of the polariton system, i.e. $E_0^{lp}=\mu$. The grand-canonical
thermal average has to be taken accordingly.

\subsection{Beliaev equations}

In analogy with the standard field theory for a single Bose field
\cite{shi98}, the $4\times 4$ matrix propagator $G({\bf
k},i\omega_n)$ obeys the Dyson-Belaev equation
\begin{equation}
G\left({\bf k},i\omega_n\right)=G^{0}\left( {\bf
k},i\omega_n\right)\left[{\bf 1}+\Sigma\left({\bf
k},i\omega_n\right) G\left({\bf k},i\omega_n\right)\right],
\label{eq:dyson_kdep}
\end{equation}
where we have introduced the matrix of the non-interacting
propagators
\begin{equation}
G^{0}\equiv \{g^{0}_{jl}({\bf
k},i\omega_n)\}_{jl}^{\chi\xi}=\delta_{\chi\xi}\delta_{jl}[(-)^{j}i\omega_n-\epsilon_{
k}^{(\xi)}+\mu]^{-1}
\end{equation}
and the $4\times 4$ self-energy matrix
\begin{equation}
\Sigma({\bf k},i\omega_n)=\left(\begin{array}{c c} \Sigma^{xx}(
{\bf k},i\omega_n) & \Sigma^{xc}( {\bf
k},i\omega_n) \\
\Sigma^{cx}({\bf k},i\omega_n) & \Sigma^{cc}({\bf k},i\omega_n)
\end{array}\right)\,. \label{eq:selfen_kdep}
\end{equation}
Within the HFP limit, the self-energy elements are independent of
frequency and read
\begin{eqnarray}
&&\Sigma_{jj}^{xx}({\bf k})=2\sum_{\bf q}\left[\bar{v}_x({\bf
k,q})
n^{xx}_{\bf q}-\bar{v}_{s}({\bf k,q})\left(n^{cx}_{\bf q}+n^{xc}_{\bf q}\right)\right], \nonumber \\
&&\Sigma_{12}^{xx}({\bf
k})=\left(\Sigma_{21}^{xx}\right)^{*}=\bar{v}_x({\bf k,0})
\Phi_{x}^{2}-2\bar{v}_{s}({\bf k,0})\Phi_{x}\Phi_{c}, \nonumber \\
&&\Sigma_{11}^{xc}({\bf
k})=\Sigma_{22}^{xc}({\bf k})=\hbar\Omega_{R}\left(1-2\sum_{\bf q} \bar{v}_s({\bf k,q})n^{xx}_{\bf q}\right), \nonumber \\
&&\Sigma_{12}^{xc}({\bf k})=\left(\Sigma_{21}^{xc}({\bf
k})\right)^{*}=-\bar{v}_{s}\Phi_{x}^{2},
\label{eq:sigmapopov_kdep}
\end{eqnarray}
while $\Sigma_{jl}^{cx}({\bf k})=\Sigma_{jl}^{xc}(-{\bf k})$ and
$\Sigma_{jl}^{cc}({\bf k})=0$.

The solutions of Eq.~(\ref{eq:dyson_kdep}) can be written
analytically in terms of the self-energy elements and the
unperturbed propagators. For example we obtain
\begin{equation}
g^{xx}_{11}({\bf p})=\frac{g^{x}_{0}({\bf p}) \left[ 1 -
g^{x}_{0}(-{\bf p}) N_D^*({\bf p}) \right]}{\left| 1 -
g^{x}_{0}({\bf p}) N_D({\bf p}) \right|^2 - \left| g^{x}_{0}({\bf
p}) N_B({\bf p}) \right|^2}\,, \label{eq:sol_dys}
\end{equation}
where ${\bf p}\equiv {\bf k},i\omega_n$,
\begin{eqnarray}
N_D({\bf p})&=&\Sigma_{11}^{xx}({\bf k}) + g_{0}^{c}({\bf p})
|\Sigma_{11}^{xc}({\bf k})|^2
 + g_{0}^{c}(-{\bf
p}) |\Sigma_{12}^{xc}({\bf k})|^{2}\,,
\nonumber \\
N_B({\bf p})&=&\Sigma_{12}^{xx}({\bf k}) + \left[g_{0}^{c}({\bf p})
+ g_{0}^{c}(-{\bf p})\right]\Sigma_{11}^{xc}({\bf k})
\Sigma_{12}^{xc}({\bf k})\,, \nonumber
\end{eqnarray}
and
\begin{equation}
g^{xx}_{21}({\bf p})=\frac{g^{x}_{0}(-{\bf p}) N^*_B({\bf
p})}{\left[ 1 - g^{x}_{0}(-{\bf p}) N^*_D({\bf p})
\right]}g^{xx}_{11}({\bf p})\,. \label{eq:sol_dys2}
\end{equation}
For each value of ${\bf k}$, the analytic continuation of each
Green's function $g_{jl}^{\chi\xi}({\bf k},z)$ shares the same
four simple poles at $z=\pm E^{lp(up)}_{{\bf k}}$, i.e.
\begin{eqnarray}
g^{xx}_{11}({\bf k},z)&=&\sum_{j=lp,up}\frac{|X^{j}_u({\bf
k})|^2}{z-E^{j}({\bf k})} + \frac{|X^{j}_v({\bf
k})|^2}{z+E^{j}({\bf k})^*} \nonumber \\
g^{xx}_{12}({\bf k},z)&=&\sum_{j=lp,up}\frac{X^{j}_u({\bf
k})^*X^{j}_v({\bf k})}{z-E^{j}({\bf k})} + \frac{X^{j}_v({\bf
k})^*X^{j}_u({\bf
k})}{z+E^{j}({\bf k})^*} \nonumber \\
g^{cc}_{11}({\bf k},z)&=&\sum_{j=lp,up}\frac{|C^{j}_u({\bf
k})|^2}{z-E^{j}({\bf k})} + \frac{|C^{j}_v({\bf
k})|^2}{z+E^{j}({\bf k})^*} \nonumber \\
g^{xc}_{11}({\bf k},z)&=&\sum_{j=lp,up}\frac{X^{j}_u({\bf
k})^*C^{j}_u({\bf k})}{z-E^{j}({\bf k})} + \frac{X^{j}_v({\bf
k})^*C^{j}_v({\bf k})}{z+E^{j}({\bf k})^*} \nonumber \,,
\label{eq:sol_Hopf}
\end{eqnarray}
and so on.\footnote{Here we write the general expression with
complex energies $E^{lp(up)}({\bf k})$. Within such a notation,
the formulas can be in principle extended to include a
phenomenological imaginary part to the energies, in order to
account for the finite radiative lifetime of polaritons.} The
poles of the propagators represent the positive and negative
Bogoliubov-Beliaev eigen-energies of the lower- and
upper-polariton modes. The residual in each pole depends on the
corresponding generalized Hopfield coefficients.

We point out that the polariton excitation modes for a given
${\bf k}$ can be also obtained by directly diagonalizing the
problem
\begin{equation}
E({\bf k})\left(\begin{array}{c}X_u \\ X_v \\ C_u \\ C_v
\end{array}\right)({\bf k})=\hat{L}_{HFB}({\bf k}) \left(\begin{array}{c}
X_u \\
X_v \\
C_u \\
C_v \end{array}\right)({\bf k})\,,
\end{equation}
with
\begin{equation}
\hat{L}_{HFP}=\left(\begin{array}{cccc}
\tilde{\epsilon}_{\bf k}^x + \Sigma_{11}^{xx} & \Sigma_{12}^{xx} & \Sigma^{xc}_{11} & \Sigma_{12}^{xc} \\
-\Sigma_{21}^{xx} & -(\tilde{\epsilon}_{\bf k}^x +\Sigma_{22}^{xx})^* & -\Sigma^{xc}_{21} & -\Sigma_{22}^{xc} \\
\Sigma^{cx}_{11} & \Sigma_{12}^{cx} & \tilde{\epsilon}_{\bf k}^c - \mu  & 0 \\
-\Sigma_{21}^{xc} & -\Sigma^{cx}_{11} & 0 & -\tilde{\epsilon}_{\bf
k}^{c*}\end{array}\right)\,,
\end{equation}
and $\tilde{\epsilon}_{\bf k}^{x(c)}=\epsilon_{\bf k}^{x(c)} -\mu$.
The components of the 4 eigenvectors ${\bf h}_j(k)\equiv (X_u, X_v,
C_u, C_v)_j(k)$ ($j=1,...,4$) are again the generalized Hopfield
coefficients corresponding to the normal $(X_u, C_u)$ and anomalous
$(X_v, C_v)$ components of the polariton field, in analogy with the
one-field HFP theory. They obey the normalization relation
\begin{equation}
|X^{j}_u|^2-|X^{j}_v|^2+|C^{j}_u|^2-|C^{j}_v|^2=1\,,
\label{eq:genHopfnorm}
\end{equation}
a condition which guarantees that the operator destroying the
lower (upper) polariton excitation with wave vector ${\bf k}$,
\begin{eqnarray}
\hat{\pi}^{j}_{\bf k} &=& X^{j}_u ({\bf k}) \hat{b}_{\bf
k} + X^{j}_v({\bf k}) \hat{b}^{\dagger}_{-\bf k} \nonumber \\
&&+ C^{j}_u({\bf k}) \hat{b}_{\bf k} + C^{j}_v({\bf k})
\hat{c}^{\dagger}_{-\bf k} \,, \label{eq:piHopf}
\end{eqnarray}
$j=lp,up$, obey Bose commutation rules. The lower (upper)
polariton one-particle operators $\hat{p}^{j}_{\bf k}$ are then
defined by
\begin{equation}
\hat{\pi}^{j}_{\bf k} = u^{j}({\bf k}) \hat{p}_{\bf k}+
v^{j}(-{\bf k})^* \hat{p}_{-\bf k}^{\dagger} \,,
\end{equation}
where the normal and anomalous polariton coefficients are given by
\begin{eqnarray}
&&u^{j}({\bf k})=\left[X^{j}_u ({\bf k}) + C^{j}_u ({\bf k})\right]^{1/2}\,, \nonumber \\
&&v^{j}({\bf k})=\left[X^{j}_v ({\bf k}) + C^{j}_v ({\bf k})\right]^{1/2}\,, \nonumber \\
&&|u^{j}({\bf k})|^2-|v^{j}({\bf k})|^2=1\,. \label{eq:quantumflpol}
\end{eqnarray}
The normal modes of excitation are thermally populated via the Bose
distribution
\begin{equation}
\bar{N}^{j}_{\bf k}\equiv \langle \hat{\pi}^{j\dagger}_{\bf k}
\hat{\pi}^{j}_{\bf k} \rangle =\frac{1}{e^{\beta E^{j}_{\bf k}
}-1}\,,
\end{equation}
while the lower- and upper-polariton one-particle densities are
given by
\begin{equation}
\tilde{n}^{j}_{\bf k}\equiv \frac{1}{A} \left[\left(|u^{j}({\bf
k})|^2+|v^{j}({\bf k})|^2\right) \bar{N}^j_{\bf k} + |v^{j}({\bf
k})|^2\right]\,. \label{eq:quantumflpol2}
\end{equation}
The first and the second term of the sum represent the thermal
and quantum fluctuations, respectively. Therefore, for a fixed
total polariton one-particle density $n_{p}$, the density of the
polariton condensate is given by
\begin{equation}
n_0 \equiv |\Phi|^2=n_{p}-\sum_{{\bf k}\neq 0}
[\tilde{n}^{lp}_{\bf k}+\tilde{n}^{up}_{\bf k}]\,.
\label{eq:condfr}
\end{equation}
From $n_0$, the exciton and the photon condensed densities are
finally obtained via Eq.~(\ref{eq:GPeq_matr2}).

Hence, for a given polariton density $n_p$ and temperature $T$, a
self-consistent solution can be obtained by solving iteratively
Eqs. (\ref{eq:GPeq_matr}), (\ref{eq:dyson_kdep}),
(\ref{eq:densgr}) and (\ref{eq:condfr}), until convergence of the
chemical potential $\mu$ and the density matrix
${n}_{\chi\xi}({\bf k})$ is reached. From this self-consistent
solution, we obtain the exciton and photon components of the
condensate fraction as well as the spectrum of collective
excitations and the one-particle populations. We point out that
the self-consistent solution must be independent on the initial
condition used in Eq.~(\ref{eq:dyson_kdep}) and
(\ref{eq:GPeq_matr}). Fast convergence of the iterative procedure
in the numerical calculations is obtained by starting from the
ideal gas solution, i.e. the solution obtained by neglecting the
two-body interactions, and considering the resulting polariton
states occupied accordingly to the Bose distribution.

\section{Results~and~discussion}
\label{sec:discuss}

For calculations we adopt parameter modeling typical GaAs based
microcavity samples \cite{balili07,deng06,deng03}, in particular
we assume a linear coupling strength
$\hbar\Omega_{R}=7~\mbox{meV}$, corresponding to 12 embedded GaAs
quantum wells, and the photon-exciton detuning
$\delta=\epsilon^{c}_0-\epsilon^{x}_0=3$~meV. Where not
differently specified, we consider a system area $A = 1000~\mu
\mbox{m}^2$ and a polariton temperature $T=10$~K. For the
interaction potentials $v_x$ and $v_s$, we use momentum-dependent
values following Rochat et al.~\cite{rochat00}.

\subsection{Spectral and thermodynamic properties}

\begin{figure}
  \includegraphics*[width=\linewidth,height=.595\linewidth]{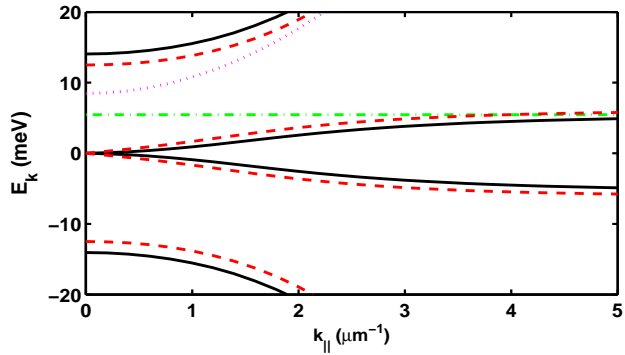}
  \caption{The dispersion of the normal modes of the system for
polariton density $n_p=5 \mu\mbox{m}^{-2}$ (solid) and $n_p=50
\mu\mbox{m}^{-2}$ (dashed). The uncoupled photon (dotted) and
exciton (dash-dotted) modes are also shown.}\label{fig:disp}
\end{figure}

\begin{figure}
  \includegraphics[width=\linewidth,height=.6\linewidth]{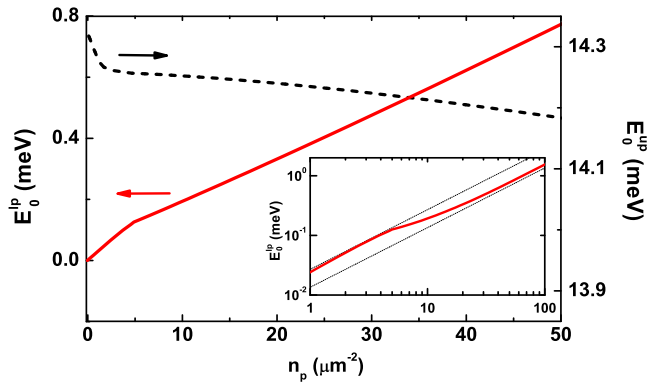}
  \caption{Lower (solid) and upper (dashed) polariton energies
at $k=0$ vs polariton density $n_p$. Inset: Double logarithmic
plot of the lower polariton energy. The thin dotted lines
highlight the two different slopes below and above the density
threshold.}\label{fig:shifts}
\end{figure}

In Fig. \ref{fig:disp} we show the energy-momentum dispersion of
the collective excitations, $\pm E^{lp}_{{\bf k}}$ and $\pm
E^{up}_{{\bf k}}$, as obtained at the critical polariton density
$n_p=5~\mu\mbox{m}^{-2}$ and far above the critical density, i.e.
$n_p=50~\mu\mbox{m}^{-2}$. The curves correspond to the positive-
and negative-weight resonances for the lower- and the
upper-polariton. We notice that, for the largest value of $n_p$,
the polariton splitting decreases, due to both the exciton
saturation, decreasing the effective exciton-photon coupling
$\Sigma^{xc}_{11}$, and the change of the exciton-photon detuning
produced by the exciton blueshift, given by $\Sigma_{11}^{xx}$.
However this variation is quantitatively small, suggesting that,
at equilibrium, the polariton structure should be robust even far
above the condensation threshold. We also mention that, close to
zero momentum, the dispersion of the lower polariton branch,
above threshold, becomes linear, giving rise to phonon-like
Bogolubov modes, as in the standard equilibrium single-field
theory \cite{pita03}.
The modification of the energy splitting between the lower and
the upper polariton branch is accurately characterized in Fig.
\ref{fig:shifts}, where the energy shifts of the two polariton
modes at ${\bf k}=0$ are plotted as a function of the density.
Exciton saturation and interactions result in a global blue-shift
of the lower polariton and a red-shift of the upper polariton. The
shifts are linear as a function of the density, but their slope
varies close to threshold. As highlighted in the inset, the slope
of the lower polariton shift changes by a factor of two across
the threshold, because the contribution of the condensed
populations ($n^{0}_{xx}$, $n^0_{xc}$) is one half the
contribution of the thermal populations ($\tilde{n}_{xx}$,
$\tilde{n}_{xc}$), as seen in Eq.~(\ref{eq:GPeq}).

\begin{figure}
  \includegraphics*[width=\linewidth,height=1.0\linewidth]{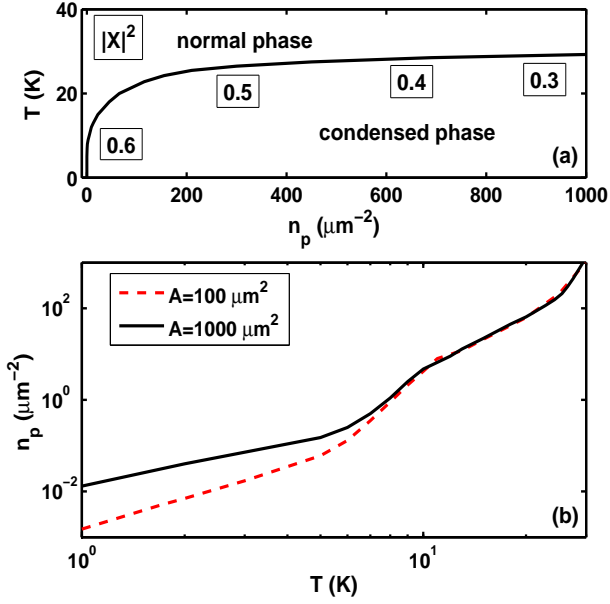}
  \caption{(a) Phase diagram of the polariton condensation, computed
for $A=1000~\mu \mbox{m}^2$. The exciton fraction in the
condensate $|X_0|^2$, along the phase boundary, is indicated in
boxes. (b) Detail of the low-$T$ region where we also display the
transition boundary computed for $A=100~\mu
\mbox{m}^2$.}\label{fig:phdiagr}
\end{figure}

We now turn to the thermodynamic properties of the system. In
Fig. \ref{fig:phdiagr}(a), we report the density-temperature BEC
phase diagram, as computed for $A=1000~\mu \mbox{m}^2$. The phase
boundary in the calculations has been set by the occurrence of a
finite fraction of polariton condensate larger than $1\%$. In the
plot, a few values of the quantity $|X_0|^2$ along the phase
boundary are indicated. This quantity represents the exciton
amount in the polariton condensate. It decreases for increasing
density, depending on the exciton saturation and the change in
detuning. For very large densities this quantity eventually
vanishes, corresponding to the crossover to a photon-laser
regime. However, for the studied GaAs model microcavity, the
variation of the exciton amount in the condensate field remains
very small up to densities far above the experimentally estimated
polariton density \cite{deng06,deng03}. This is basically due to
the positive cavity detuning, sufficiently large to be robust to
the exciton nonlinear energy shift. On the other hand, due to the
same feature and to the flat energy dispersion of the exciton-like
lower polariton states, a large population can be accommodated in
the excited state when the temperature exceeds $25$~K, thus
dramatically increasing the BEC transition density, eventually
leading to the direct occurrence of photon-lasing. In particular,
for this system, we predict that equilibrium polariton BEC is
impossible for temperatures larger than $30$~K. In Fig.
\ref{fig:phdiagr}(b), we show a detail of the low-$T$ region of
the phase diagram, computed for two different system areas
$A=100~\mu \mbox{m}^2$ and $A=1000~\mu \mbox{m}^2$. In a
homogeneous two-dimensional system, in the limit of infinite
size, a true condensate cannot exist due to the divergence of
low-energy thermal fluctuations. The transition to a superfluid
state is instead expected, giving rise to the
Berezinski-Kosterlitz-Thouless crossover with spontaneous
unbinding of vortices. The divergence of the condensate
fluctuations has however a logarithmic dependence on the system
size. Fig. \ref{fig:phdiagr}(b) shows this behaviour as a slow
increase of the critical density for increasing $A$.
Quantitatively, the critical density varies by no more than a
factor 2 at $T=1$~K, for the two considered values of the system
area. This difference becomes even smaller for larger
temperatures. The predicted dependence on the system size could
be experimentally verified only in samples with improved
interface quality and manifesting thermalization at very low
polariton temperature \cite{sarchi07b}.

\section{Linear response in the HFP limit}

Within the present theory, it is easy to compute the density
fluctuation of the exciton or the photon field produced by a
perturbation acting on either of them. This quantity is
particularly interesting because as we suggest later, it might be
used to study the nature of the polariton condensate via fully
optical or mechanical perturbations.

We consider the time dependent perturbation
\begin{equation}
\hat{H}_{pert}(t)=\int d{\bf r} \hat{n}^{\chi\chi}({\bf
r},t)V_{ext}({\bf r},t)\,, \label{eq:hpert}
\end{equation}
driven by the external potential $V_{ext}({\bf r},t)$ affecting
the field $\hat{\Psi}_{\chi}({\bf r},t)$ ($\chi=x,c$). This
perturbation results in a density fluctuation
\begin{equation}
\delta n^{\xi\xi}({\bf r},t) \equiv n^{\xi\xi}({\bf r},t) -
n_{eq}^{\xi\xi}({\bf r},t)\,,
\end{equation}
($\xi=x,c$) around the equilibrium value $n^{eq}_{\xi\xi}({\bf
r},t)$. Within the linear response limit, this fluctuation is
given by \cite{pita03,fetter71} \footnote{We are only considering
the response of the exciton and/or the photon density to a
perturbation affecting one of the two species. For this reason we
are not considering expressions involving the non diagonal terms
of the density operator $\hat{n}^{xc}$.}
\begin{equation}
\delta n^{\xi\chi}({\bf r},t)=\frac{i}{\hbar}\int_{-\infty}^t dt'
\int d{\bf r'} C_{\xi\chi}({\bf r'},t';{\bf r},t) V_{ext}({\bf
r'},t')\,, \label{eq:densresp}
\end{equation}
where
\begin{equation}
C_{\xi\chi}({\bf r'},t';{\bf
r},t)=\langle\left[\hat{n}^{\chi\chi}({\bf
r'},t'),\hat{n}^{\xi\xi}({\bf
r},t)\right]\rangle\,.\label{eq:denscomm}
\end{equation}
For a spatially uniform system in steady state, the density
commutator Eq. (\ref{eq:denscomm}) only depends on ${\bf r-r'}$
and $t-t'$. Then, by Fourier transforming Eq.
(\ref{eq:densresp}), we get the expression
\begin{eqnarray}
\delta n^{\xi\chi}({\bf k},\omega)=D^{r}_{\xi\chi}({\bf
k},\omega)V_{ext}({\bf k},\omega)\,, \label{eq:densresp2}
\end{eqnarray}
where
\begin{equation}
D^{r}_{\xi\chi}({\bf k},\omega)=-\frac{1}{\hbar}\int
\frac{d\omega'}{2\pi}\frac{C_{\xi\chi}({\bf
k},\omega-\omega')}{\omega'+i0^+}
\end{equation}
is the retarded density-density correlation \cite{fetter71}, and
$C_{\xi\chi}({\bf k},\omega)$ is the Fourier transform of
$C_{\xi\chi}({\bf r},t)$.

Using the framework developed in Section \ref{sec:theory}, we
write the real-time density operators $\hat{n}^{\chi\chi}({\bf
r},t)$ in the polariton basis, via
Eqs.~(\ref{eq:fieldopb}-\ref{eq:fieldopc}) and by inverting the
transformation Eq. (\ref{eq:piHopf})
\begin{equation}
\hat{O}_{\chi}({\bf k},t)=\sum_{j=lp,up}\Pi^j_{\chi}({\bf
k})e^{-i\frac{E_{\bf k}^{j}}{\hbar} t} \hat{\pi}^{j}_{\bf
k}+\Upsilon^j_{\chi}({\bf k}) e^{i\frac{E_{\bf k}^{j}}{\hbar} t}
\hat{\pi}^{j\dagger}_{-\bf k}\,.
\end{equation}
Here, the factors $\Pi^j_{\chi}({\bf k})$ and
$\Upsilon^j_{\chi}({\bf k})$ define the components of the exciton
(photon) field on the forward and backward propagating lower and
upper polariton eigenmodes.  Within the HFP limit, the density
commutator consists of three contributions
\cite{minguzzi97}\footnote{In the HFP approximation, the coupling
between the density fluctuations of the condensate and the non
condensate is neglected, because the HFP ground state is defined
as the vacuum of quasiparticles. For a trapped gas of atoms, this
approximation is found to be in qualitative agreement with
experiments, although, close to $T_c$, deviations are reported. A
better quantitative agreement has been obtained beyond the HFP
limit, by means of Random Phase Approximation techniques
\cite{minguzzi97,minguzzi04}. For our present purposes, however,
the HFP limit is an adequate approximation.}
\begin{eqnarray}
C_{\xi\chi}({\bf r},t)&=&^0C_{\xi\chi}^{I}({\bf
r},t)+^0C_{\xi\chi}^{II}({\bf
r},t)+\tilde{C}_{\xi\chi}({\bf r},t) \\
&=&\left(\Phi^*_{\xi}(0)\Phi_{\chi}(t)\langle
\left[\tilde{\psi}_{\xi}({\bf
0},0),\tilde{\psi}_{\chi}^{\dagger}({\bf
r},t)\right]\rangle-h.c.\right)\nonumber \\
&+&\left(\Phi^*_{\xi}(0)\Phi^*_{\chi}(t)\langle
\left[\tilde{\psi}_{\xi}({\bf 0},0),\tilde{\psi}_{\chi}({\bf
r},t)\right]\rangle-h.c.\right)\nonumber \\
&+&\langle \left[\tilde{\psi}^{\dagger}_{\xi}({\bf
0},0)\tilde{\psi}_{\xi}({\bf
0},0),\tilde{\psi}_{\chi}^{\dagger}({\bf
r},t)\tilde{\psi}_{\chi}({\bf r},t)\right]\rangle \nonumber\,,
\end{eqnarray}
the first two terms arising from the presence of the condensate
fields. Correspondingly, the retarded density-density correlation
can be written as the sum of three terms
\begin{equation}
D^{r}_{\xi\chi}({\bf k},\omega)=
\hspace{.05cm}^{0}D^{I}_{\xi\chi}({\bf
k},\omega)+\hspace{.05cm}^0D^{II}_{\xi\chi}({\bf
k},\omega)+\tilde{D}_{\xi\chi}({\bf k},\omega)\,.
\end{equation}
The first two terms survive only in the presence of a condensate
while the third one only depends on the thermal population
(however it is affected by the modification of the one-particle
spectrum induced by the condensate). In detail, the first term
describes the excitation of particles out of the condensate and
it is given by
\begin{equation}
^0D^{I}_{\xi\chi}({\bf
k},\omega)=\sum_{j=lp,up}\left[N^j_{\xi\chi}({\bf
k},\omega)+N^j_{\xi\chi}({\bf k},-\omega)^*\right]\,,
\end{equation}
with
\begin{equation}
N^j_{\xi\chi}({\bf
k},\omega)=\frac{\Phi_{\xi}\Phi^*_{\chi}\Pi_{\xi}^{j*}({\bf
k})\Pi_{\chi}^j({\bf
k})+\Phi_{\xi}^*\Phi_{\chi}\Upsilon_{\xi}^{j*}({\bf
k})\Upsilon_{\chi}^j({\bf k})}{\hbar\omega-E_{\bf k}^j+i0^+}\,.
\label{eq:respterm1}
\end{equation}
The second term describes the de-excitation of the thermal
population into the condensate and it is given by
\begin{equation}
^0D^{II}_{\xi\chi}({\bf
k},\omega)=\sum_{j=lp,up}\left[A^j_{\xi\chi}({\bf
k},\omega)+A^j_{\xi\chi}({\bf k},-\omega)^*\right]\,,
\end{equation}
with
\begin{equation}
A^j_{\xi\chi}({\bf
k},\omega)=\frac{\Phi_{\xi}\Phi_{\chi}\Pi_{\xi}^{j*}({\bf
k})\Upsilon^j_{\chi}({\bf
k})+\Phi_{\xi}^*\Phi^*_{\chi}\Upsilon_{\xi}^{j*}({\bf
k})\Pi^j_{\chi}({\bf k})}{\hbar\omega-E_{\bf
k}^j+i0^+}\,.\label{eq:respterm2}
\end{equation}
The third term describes the oscillations of the thermal
population and it is given by
\begin{equation}
\tilde{D}_{\xi\chi}({\bf k},\omega)=\sum_{j=lp,up}
\left[T^j_{\xi\chi}({\bf k},\omega)+T^j_{\xi\chi}({\bf
k},-\omega)^*\right]\,,
\end{equation}
with
\begin{eqnarray}
T^j_{\xi\chi}({\bf k},\omega)&=&\sum_{l,{\bf
q}}\left[\frac{F^{jl}_{\xi\chi}({\bf k,q})\left(\bar{N}^l_{\bf
q}-\bar{N}^j_{\bf q-k}\right)}{\hbar\omega+E_{\bf q-k}^j-E_{\bf
q}^l+i0^+}\right. \nonumber \\
&&+\left.\frac{R^{jl}_{\xi\chi}({\bf k,q})\left(1+\bar{N}^l_{\bf
q}+\bar{N}^j_{\bf q-k}\right)}{\hbar\omega+E_{\bf q-k}^j+E_{\bf
q}^l+i0^+}\right]\,
\end{eqnarray}
and
\begin{eqnarray}
F^{jl}_{\xi\chi}({\bf k,q})&=&\Pi^j_{\xi}({\bf
q-k})\Pi_{\chi}^{j*}({\bf q-k})\Pi_{\xi}^{l*}({\bf
q})\Pi_{\chi}^l({\bf q})\nonumber \\
&+&\Pi_{\xi}^j({\bf q-k})\Upsilon_{\chi}^{j*}({\bf
q-k})\Pi_{\xi}^{l*}({\bf q})\Upsilon_{\chi}^{l}({\bf q})\,, \\
R^{jl}_{\xi\chi}({\bf k,q})&=&\Pi^j_{\xi}({\bf
q-k})\Pi_{\chi}^{j*}({\bf q-k})\Upsilon_{\xi}^{l}({\bf
q})\Upsilon_{\chi}^{l*}({\bf q})\nonumber \\
&+&\Pi_{\xi}^j({\bf q-k})\Upsilon_{\chi}^{j*}({\bf
q-k})\Upsilon_{\xi}^l({\bf q})\Pi_{\chi}^{l*}({\bf q})\,.
\end{eqnarray}
We use these equations to study the fluctuation $\delta n_{cc}$ of
the photon density, directly related to the photoluminescence
measured in experiments, produced by an optical (affecting the
photon field) or mechanical (affecting the exciton field)
perturbation. We take, as external potential, a plane wave with
wave vector ${\bf k_{ext}}$, delta-pulsed in time, i.e.
$V_{ext}({\bf r},t)=V_0 e^{i{\bf k}_{ext} \cdot {\bf r}}
\delta(t-t_0)$. In this case, from Eq. (\ref{eq:densresp2}), we
see that $\delta n^{\xi\chi}({\bf k},\omega)=V_0
D^{r}_{\xi\chi}({\bf k},\omega)\delta({\bf k-k}_{ext})$, i.e. the
response is diagonal in ${\bf k}$ and is simply proportional to
the correlation $D^r$. The imaginary part of $D^r$ describes the
energy transfer to the system and thus it is the most relevant
function.
\begin{widetext}
\begin{center}
\begin{figure}
  \includegraphics[width=.7\textwidth]{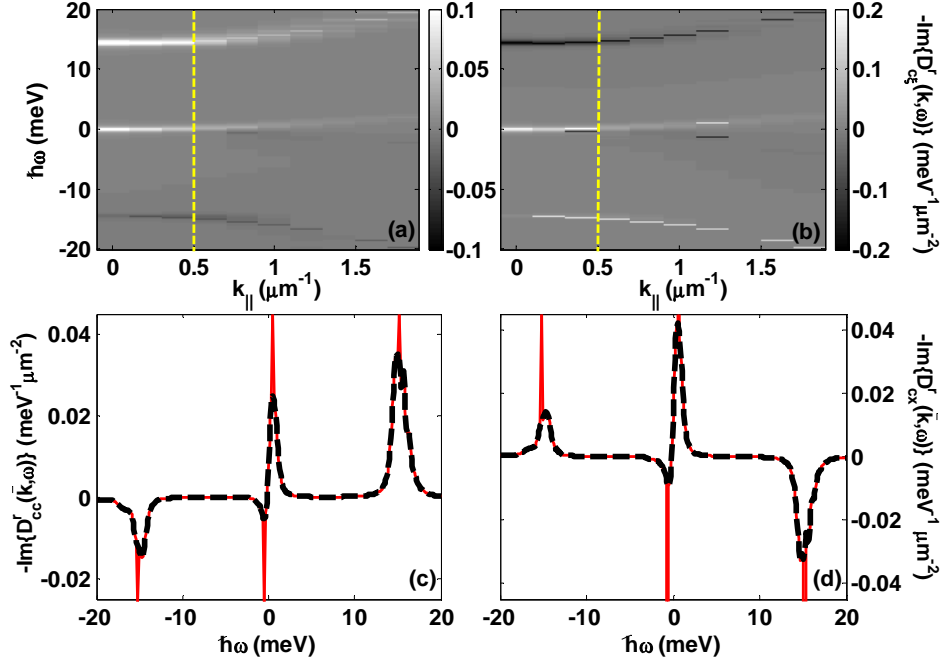}%
  \caption[]{Imaginary part of the retarded density-density correlations $D^r_{cc}({\bf
  k},\omega)$ (panel a and c)
  and $D^r_{cx}({\bf k},\omega)$ (b and d), as computed for $T=10$~K and $n_p=5~\mu\mbox{m}^{-2}$.
  In panels (a) and (b) we display the $k-\omega$ dependence of the correlation in grey tones.
  In panels (c) and (d) the same quantities are plotted as a function of the energy $\hbar\omega$ and at
  the wave vector $\bar{k}=0.5~\mu\mbox{m}^{-1}$. In panels (c)-(d), the dashed line represents
  the contribution $\tilde{D}^r_{\xi\chi}({\bf k},\omega)$ arising from the oscillations of the non condensate.
  Delta peaks arise from the oscillation of the condensate. At high
  energy (i.e. at the upper polariton energy), the photon-photon response corresponds to absorption at positive energy and gain
  at negative energy, while the photon-exciton response has the opposite behavior.}\label{fig:linresp}
\end{figure}
\end{center}
\end{widetext}
The resulting quantities $-\mbox{Im}\{D^{r}_{c\chi}({\bf
k},\omega)\}$ with $\chi=c,x$ are shown in
Fig.~\ref{fig:linresp}. We display the results for the
photon-photon (panels (a)-(c)) and photon-exciton (panels
(b)-(d)) correlation in the condensed regime. Poles at negative
energy are present for both quantities, due to the spectrum
modification induced by condensation. While the condensate
contribution defines collective modes with infinite lifetime
(delta-peaks), the contribution from the thermal population is
responsible for a finite linewidth, as can be argued from the
equations above. The surprising result of this analysis is however
the unusual behavior manifested by the photon-exciton response at
high energy. In this case, the collective mode corresponds to
gain at positive energy and to absorption at negative energy.
Conversely, for the photon-photon response, we expect to observe
the opposite behavior. This feature is due to the phases of the
Hopfield coefficients. As shown by Eqs.
(\ref{eq:respterm1},\ref{eq:respterm2}), the opposite nature of
the collective modes generated by optical or mechanical
perturbations would be observed in experiments only if an exciton
coherent field $\Phi_x$ is present. Furthermore, the relative
amplitude of the two responses is proportional to the fraction of
coherence of each field, giving direct access to the amount of
exciton condensate. In this respect, we point out that, although
any kind of perturbation would affect simultaneously both the
exciton and the photon field \cite{hvam06}, the geometry of the
system can be chosen in such a way that the effect of the
perturbation on one of the two fields be dominant (see for
example the static perturbation affecting the exciton field used
by Balili \emph{et al.} \cite{balili07}). We thus suggest that an
ideal tool to observe these features, would be a pump and probe
experiment where the probe be in turn optical or mechanical (for
example produced by a coherent acoustic waves \cite{hvam06}).

\section{Conclusions}

We generalized the HFP theory to the case of two coupled Bose
fields at equilibrium. The theory allows modeling the BEC of
microcavity polaritons in very close analogy with the BEC of a
weakly interacting gas \cite{fetter71,pita03}. In particular we
treat simultaneously both the linear exciton-photon coupling and
interactions. We account for the presence of a non condensed
population as well. Within this description we are able to
predict the modification of the spectrum and the thermodynamic
properties for increasing density. Since the theory allows to
describe simultaneously the properties of the polariton, the
photon and the exciton fields, it allows the understanding of
typical optical measurements. In particular, our analysis
supports the interpretation of the recent experimental findings
\cite{kasprzak06,balili07,deng06} in terms of BEC of a trapped
gas. We have applied the theory to compute the density-density
response of a polariton gas to an external perturbation. This
quantity can be characterized in pump and probe experiments
realized with an optical or mechanical probe. In particular, we
predict that the upper polariton energy collective modes
generated in the two cases have a response of opposite sign. We
suggest that the observation of this feature would be a proof of
the presence of an exciton condensate and would answer the long
standing question about the connection between polariton BEC and
a laser phenomenon.

We acknowledge financial support from the Swiss National
Foundation through project N. PP002-110640.

\bibliography{sarchibib}

\end{document}